\documentclass[12pt]{article}
\usepackage{amssymb}
\usepackage{amsmath}
\usepackage{graphicx,psfrag,epsf}
\usepackage{enumerate}
\usepackage{natbib}
\usepackage{tabularx}
\usepackage{url} 
\usepackage{tikz}
\usetikzlibrary{calc,arrows,positioning,decorations.pathreplacing}
\usepackage{bm}
\usepackage{float}
\usepackage{multirow}
\usepackage{threeparttable}
\usepackage{authblk}
\usepackage{booktabs}
\usepackage{dsfont}
\usepackage{tikz-network}

\newtheorem{assum}{Assumption}

\usepackage{color,verbatim}

\begin{document}

\def\spacingset#1{\renewcommand{\baselinestretch}%
{#1}\small\normalsize} \spacingset{1}



  \title{\bf General Regression Methods for Respondent-Driven Sampling Data}
\author[1]{Mamadou Yauck\thanks{
  E-mail: \textit{mamadou.yauck@mcgill.ca}}}
\author[1]{Erica E. M. Moodie}
\author[1]{Herak Apelian}
\author[1]{Alain Fourmigue}
\author[3]{Daniel Grace}
\author[4]{Trevor Hart}
\author[2]{Gilles Lambert}
\author[1]{Joseph Cox}
\affil[1]{McGill University, Montreal, Québec, Canada}
\affil[2]{Institut national de santé publique du Québec, Montreal, Québec, Canada}
\affil[3]{Dalla Lana School of Public Health, University of Toronto}
\affil[4]{ Ryerson University, Toronto, Ontario, Canada}
  \maketitle


\bigskip
\begin{abstract}
Respondent-Driven Sampling (RDS) is a variant of link-tracing sampling techniques that aim to recruit hard-to-reach populations by leveraging individuals' social relationships. As such, an RDS sample has a graphical component which represents a partially observed network of unknown structure. Moreover, it is common to observe \textit{homophily}, or the tendency to form connections with individuals who share similar traits. Currently, there is a lack of principled guidance on multivariate modeling strategies for RDS to address homophilic covariates and the dependence between observations within the network. In this work,  we propose a methodology for general regression techniques using RDS data. This is used to study the socio-demographic predictors of HIV treatment optimism (about the value of antiretroviral therapy) among gay, bisexual and other men who have sex with men, recruited into an RDS study in Montreal, Canada.
\end{abstract}

\noindent%
{\it Keywords: Hidden population sampling; identification; design weights; homophily; simultaneous autoregressive models; peer effects; social networks.}
\vfill

\newpage
\spacingset{1.45} 
\section{Introduction}
\label{sec:intro}

	\textit{Respondent-Driven Sampling} (RDS) is a network-based sampling technique that leverages social relationships to recruit individuals of hard-to-reach populations into research studies \citep{Hec97}. The RDS process, which proceeds through recruitment \textit{waves}, starts with the selection of initial \textit{seed} participants who, after being interviewed, receive a fixed number of \textit{coupons} to distribute among their peers. RDS offers many advantages over existing network-based sampling methods. Through many waves of recruitment, the process samples farther from the initial recruits, which should ensure greater representativeness and hence generalizability of the sample. This is because seeds typically represent a convenience sample, even if thoughtfully  chosen with the view to optimizing representation of their social spheres. Moreover, RDS reduces the privacy concerns that are associated with the identification of participants' social networks or the community population that could occur in a more traditional study that would aim to enumerate the members of the target population by relying on members to recruit their peers into the study.
	
	An RDS sample has a graphical structure, which is typically a partially observed social network of recruited individuals with an unknown underlying dependence structure in which it is common to observe a tendency for individuals with similar traits to share social ties, a feature termed \textit{homophily}. Moreover, the RDS process  is not one that is purely random, but rather some individuals are more likely to be selected into the sample than others. An assumed underlying principle in RDS is that the probability of an individual being recruited depends on the size of their personal network of social contacts (\citealt{Gile11,Hec97}). However, the true RDS sampling design is unknown, warranting inferential methods that rely on approximations to the true RDS process to estimate sampling weights.

	As highlighted in \cite{Gile18}, the current literature of RDS data lacks principled approaches to multivariable modeling. This is reflected in the variety of analytic approaches taken in the applied literature. Some studies have treated RDS data as though collected by random sampling and applied ANOVA, linear and logistic regressions without any adjustment for RDS sampling weights \citep{Ram13}. Others have included RDS weights in regression models, relying on the typical RDS assumption that some individuals are more likely to be recruited into the sample than others, while ignoring the dependence between observations within the RDS network \citep{Johnston2010TheAO}. In yet another approach, \cite{Rho15} included seeds as random effects to adjust for the dependence within recruitment chains but ignored RDS weights. A mixed effects model including random effects on features such as seeds and recruiters to account for the dependence, and using weights at different levels of clustering when appropriate, has been proposed by \cite{Spi09}. The author further proposed to model social effects driven by homophily 
by including a parameter to account for possible interactions between recruiters and recruits' values of homophilic covariates. This approach was presented as a general guidance for RDS regression; however no theoretical details or practical (simulation) demonstrations of the performance of the proposed methodology were provided.
	
Thus, while there are well-developed strategies for estimating means and prevalences from RDS studies, best practices for regression modeling remain poorly characterized. And yet, understanding dependence between variables is often a primary goal in epidemiologic research. Take for example the question of whether socio-demographic characteristics can predict optimism about the value of antiretroviral therapy, either as a pre-exposure prophylaxis (PrEP) or post-infection treatment, in a population of gay, bisexual and other men who have sex with men (GBM). There have been suggestions that younger people (aged less than 35) were less likely to have optimism, while people with lower annual income (less than \$20,000) were more likely to have optimism \citep{levy2017longitudinal,craib2002hiv}, which could potentially mitigate the effectiveness of HIV preventive measures.
The Engage study, which is an RDS study conducted in Montreal, Toronto and Vancouver, provides a unique opportunity to study this question in a large sample of the GBM community -- but doing so requires appropriate modeling strategies.

	One of the most challenging issues of multivariate modeling for RDS is one of missing data. In fact, the observed data reveal partial information about the full RDS network in which all connections between recruited individuals are reported (see \citealt{Weeks02, Mosher2015AQA} for a rare example of an RDS study in which those traditionally missing connections are reported). This problem is fundamentally design-based \citep{Crawford17}. In this case, when conducting inference about homophily-driven effects and/or network-induced correlation structures, different full data distributions give rise to the same distribution for the observed data. This lack of identification has been thoroughly discussed in Yauck et al.~(2020b). A crucial implication for the validity of inferential procedures is that an infinite number of observations will not yield a perfect knowledge of the parameters for homophily-driven effects or/and network dependence unless the full RDS network is observed. 

	The paper is organized as follows. In Section \ref{sec:graph}, we provide a brief background to respondent-driven sampling, and define the resulting network structure of an RDS sample where social connections can be viewed as exhibiting a correlation structure that is analogous to a spatial pattern (where the ``distance'' metric is the number of social separations between individuals). In Section \ref{sec:methodo}, we propose a generalized mixed effects model, with homophily-driven effects to deal with homophilic covariates, and with spatial random effects to model the dependence between outcomes within the network. We briefly discuss the issue of identification when the full network of recruited individuals is only partially observed by design, and the inclusion of RDS weights to account for the non-random sampling of the target population when recruited individuals (accurately) report on their personal network sizes. The validity of the proposed methodology is investigated in simulations presented in Section \ref{sec:simulations}. In Section \ref{sec: casestudy}, we analyze the Engage data collected in Montreal to investigate the relationship between HIV treatment optimism and socio-demographic characteristics, providing reliable parameter estimates and appropriate standard errors via our proposed approach. We conclude in Section \ref{sec:discussion} with a discussion of the approach and future considerations.

\section{A brief review of RDS}\label{sec:graph}

In this section, we briefly review the assumptions needed for an RDS design, and graphically display an example of the resulting observed network structure -- which is a partial view of the underlying network structure.
Suppose an infinite population in which individuals are connected by social ties. We define this as the population network and state the following:
\begin{assum}{\textbf{(The population network).}}\label{assum:subgraph}
The population network represents an infinite number of non-overlapping clusters of finite sizes.
\end{assum}
In other words, the population is clustered, with individuals partitioned into well-defined clusters. Note that in much of the RDS literature, the population is assumed to form a connected network, with no disjoint clusters. We believe that to be an overly restrictive and unrealistic assumption. For example, the Colorado Springs Project 90 study \citep{klovdahl1994social} revealed a real-world social network of 125 connected, disjoint clusters.

Now, consider an RDS process operating across social connections of the population network.

\begin{figure}[H]
\begin{center}
\includegraphics[scale=.65]{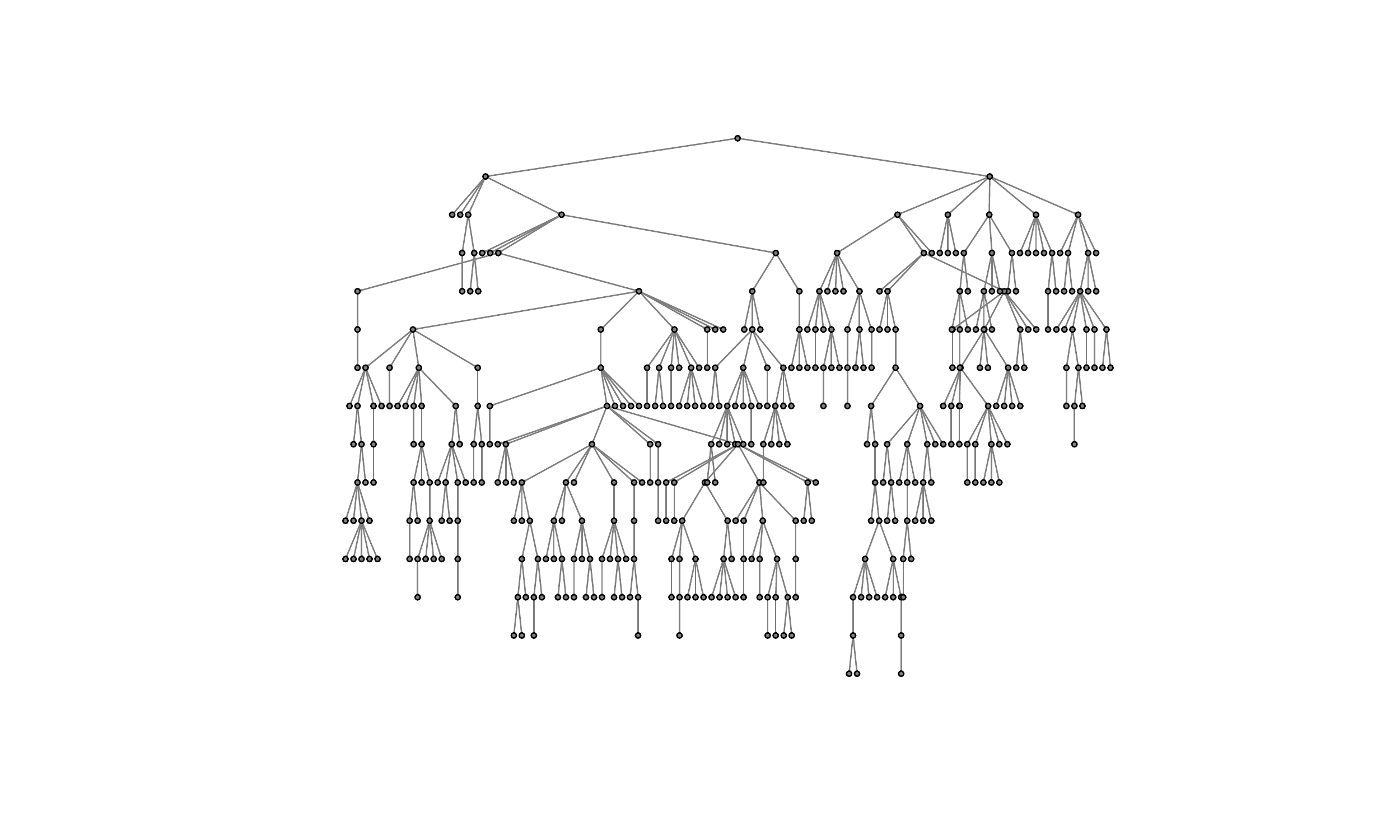}
\end{center}
\vspace*{-3cm}
\caption{Representation of the largest single tree of 412 recruits from the Engage recruitment tree of $n=1179$ gay, bisexual and  men who have sex with men in Montreal, 2018. Individuals are aligned by wave of recruitment.}
\label{fig:engagenet}
\end{figure}

\begin{assum}{\textbf{(The RDS recruitment).}}\label{assum:recruit}
The recruitment process takes place within a subset of clusters of the network and progresses across individuals' social connections.
\end{assum}
This assumption implies that the RDS sampling process can be characterized as a two-stage sampling design in which seeds and then, subsequently, additional individuals are selected from non-overlapping clusters.
 \begin{assum}{\textbf{(No multiple recruitments).}}\label{assum:once}
No individual can be recruited more than once into the study.
\end{assum}
Once again, this is not a typical RDS assumption. Previous work on theory for RDS estimators of means \citep{Vol08} has assumed that sampling take place with replacement, and yet in practice this does not occur. We therefore dispense with that unrealistic assumption.
The above three assumptions imply that the observed RDS network can be represented as a finite set of non-overlapping trees. For practical purposes, consider the Engage study in Montreal. The RDS recruitment consisted of three main steps.
\begin{itemize}
\item[Step 1.]  Sampling started off with the purposeful selection of a first group of 27 GBM, the seed participants. Seeds were selected to be representative with respect to the diversity of the GBM community based on a community mapping exercise. The seeds were invited to a community-based survey site to complete a questionnaire and to undergo testing for sexually transmitted and bloodborne infections. Seeds who successfully completed the study received a (monetary) remuneration known as primary incentive.  This is wave zero of recruitment.
\item[Step 2.] Successful seed participants were each given six uniquely identified coupons, and asked to recruit their GBM peers into the study; the social ties between a recruiter and any new participants recruited were then known to the study through the coupon, and recorded in the study database. Successful recruiters received a secondary (monetary) incentive for each peer that they recruited.
\item[Step 3.] The process continued through successive waves until the desired sample size was reached. 
\end{itemize}
Figure \ref{fig:engagenet} illustrates the largest single cluster of the RDS recruitment tree from the Engage study of GBM in Montreal.

\section{Methodology}\label{sec:methodo}

In this section, we jointly model homophily-driven effects and the dependence between outcomes from the clusters of the unobserved population network. This allows us to view the fitting of the assumed model to the observed RDS data as a missing data problem. The resulting identification issue is discussed in Section \ref{sec:designid}. Common strategies to account for the non-random sampling of the  population and the question of whether to weight the model are discussed in Section \ref{sec:RDSweights}.

\subsection{Underlying, data-generating model and assumptions}\label{sec:modelbased}
Let $y_{ij}$ be the outcome on the $j$th individual of the $i$th cluster, $j=1,\dots,N_i$, where $N_i$ is the size of the $i$th cluster, and $i=1,\dots, m$. Let $x_{ij}$ be the value of the covariate for the $j$th individual of the $i$th cluster, and $\mathbf{x}_{i}$ the vector of covariates for all individuals in the $i$th cluster. We assume that $\{y_{ij}, x_{ij}; i=1,\dots, m; j=1, \dots, N_i\}$ is the realization of a random sample whose distribution is identical to that of the superpopulation of clusters defined in Section \ref{sec:graph}, so that any inference based on the sample pertains to the parameters of the infinite population from which the sample is drawn. Inspired by \cite{Manski03}, we assume the underlying relationship between the outcome and covariates in the population is characterized by a generalized linear mixed model in which $\bm{\delta}_i=\left(\delta_{i1},\dots,\delta_{iN_i}\right)$ is a vector of random effects for the $i$th cluster, $\mu_{ij}=\mbox{E}\left(y_{ij}|\mathbf{x}_{i},\delta_{ij}\right)$, and
\begin{equation}\label{GLMmodel}
g\left(\mu_{ij}\right)=\beta_0+\beta_1 x_{ij}+\gamma \frac{1}{n_{ij}}\sum_{k\sim j}x_{ik}+\delta_{ij},
\end{equation}
where $g(.)$ is a (monotonic) function of the mean, $k \sim j$ represents the set of individuals who share ties with the $j$th individual, $n_{ij}$ is the number of social connections that the $j$th individual of the $i$th cluster shares with other individuals within the same cluster, or \textit{degree}. We further assume that $\bm{\delta}_i \sim N(\bm{0},\bm{\Sigma}_i)$, with $\mbox{cov}\left(\bm{\delta}_i,\bm{\delta}_{j}\right)=\bm{0}$ for $i\neq j$.
The parameter $\gamma$ measures homophily-driven effects, or the influence of peers' characteristics on the outcome of an individual. In this model, the parameters $\beta_0$ and (the potentially vector-valued parameter) $\beta_1$ are of primary interest.

Now let $\mathbf{S}^{(i)}=(s^{(i)}_{jk})$ be a (\textit{neighborhood}) matrix representing social ties in the $i$th cluster such that $s^{(i)}_{jk}=1$ if individual $j$ and individual $k$ share a tie and $s^{(i)}_{jk}=0$ otherwise, with $s^{(i)}_{jj}=0$, and $\mathbf{S}=\mbox{diag}(\mathbf{S}^{(i)})$.
We assume a Simultaneous Autoregressive (SAR) model (\citealt{Whittle54, cressie1993statistics}) for the vector of random effects $\bm{\delta}_i$:
\begin{equation}\label{SAR}
\bm{\delta}_i=\rho\mathbf{S}^{(i)}\bm{\delta}_i+\bm{u}_i,
\end{equation}
where $\rho$ represents the strength of the dependence within the network, and $\bm{u}_i \sim N(\bm{0},\sigma^2\bm{I}_{N_i})$. Given $\mathbf{W}_i=(\bm{I}_{N_i}-\rho\mathbf{S}^{(i)})^{-1}$ exists, the covariance of $\bm{\delta}_i$, $\bm{\Sigma}_i$, can be written as 
\begin{equation}\label{Sigmaneighbour}
\bm{\Sigma}_i(\sigma^2,\rho)=\sigma^2\mathbf{W}_i\mathbf{W}_i^{\top}.
\end{equation}
The SAR correlation matrix is such that outcomes from \textit{neighboring} (i.e.~socially connected) individuals are more correlated than outcomes from non-neighbors. Other correlation models for $\bm{\delta}_i$ with such properties include Conditional Autoregressive (CAR) models, which belong in the same class of areal models as SAR models \citep{banerjee2003hierarchical}, and models which assume a correlation function that depends on the ``distance'' between observations \citep{f2007methods}.

\subsection{Identification and the validity of inference}\label{sec:designid}
Consider the observed data from RDS $\bm{\mathcal{D}}_T=(y_{ij}, x_{ij}; i=1,\dots,m, j=1,\dots, n_i)$, where $n_i$ is the number of recruits belonging in the $i$th cluster. Let $\bm{S}_T$ represents the observed neighborhood matrix for the RDS recruitment tree. When data are collected under traditional RDS designs, the complete information on recruited inidividuals is only partially observed through $\{\mathbf{S}_T, \bm{\mathcal{D}}_T\}$. Yauck et al. (2020b) showed that, in the presence of homophily-driven effects and/or when the dependence within the network is modeled using network-induced correlation structures, traditional RDS studies suffer from the lack of identification, which arises when different full data distributions give rise to the same distribution for the observed data. This has two major implications regarding the validity of inferential procedures for model (\ref{GLMmodel}). First, an infinite number of observations will not provide a perfect knowledge of the homophily-driven effects and network-induced structure parameters unless the full RDS network is observed. Further, valid inference about those parameters can be drawn only when the recruitment tree is identical to the unobserved RDS network. Thus, fitting model (\ref{GLMmodel}) to the observed RDS data $\{\mathbf{S}_T, \bm{\mathcal{D}}_T\}$ might be an ineffective strategy.

	Now, consider the modeling of homophily-driven effects in (\ref{GLMmodel}). \cite{Spi09} recommended the inclusion of a regression parameter (the equivalent of $\gamma$) to account for a possible effect of the recruiter's value of the homophilic covariate on the outcome of the recruit. \cite{Ave19} showed empirically that ignoring that effect induces a minimal loss of precision but does not add any bias to the estimator for $\beta_1$ when fitting the model to the observed RDS data. This is encouraging since the  homophily-driven effects ($\gamma$), in model (\ref{GLMmodel}), cannot be consistently estimated given $\{\mathbf{S}_T, \bm{\mathcal{D}}_T\}$. Following these results, Yauck et al.~(2020b) proposed fitting a regression model \textit{without} homophily-driven effects to the observed data as a way of minimizing the risk of performing misleading inference when the observed RDS network is incomplete. The accuracy and precision of $\hat \beta_1$, and the coverage of 95\% confidence interval for $\beta_1$ when $\gamma$ is omitted from the analytic model will be investigated via simulations in Section \ref{sec:simulations}.
	
	Further, consider the SAR model for the random effects $\bm{\delta}_i$. Due to the lack of identification, the parameters for the induced correlation structure, which is a function of the neighborhood matrix of social connections, is inestimable given the observed data (Yauck et al., 2020b): it is not possible to adequately model $\bm{\Sigma}_i$ with incomplete information on the social ties within the observed recruitment tree. Other network-induced correlation structures such as the autoregressive, the `RDS-tree' \citep{Beckette018272} and the Toepliz, although suitable for the branching structure of the recruitment tree, also fail to adequately capture the correlation structure for the aforementioned reason. In Section \ref{sec:simulations}, we consider an alternative class of correlation models for which the dependence within the $i$th tree is induced by a cluster-specific random effect $\bm{\delta}_i=\delta_i,\,i=1,\dots,m$; clustering is assumed at the seed level and at the recruiter level \citep{Spi09}. The accuracy and precision of $\hat \beta_1$, and the coverage of 95\% confidence interval for $\beta_1$ in this case of model misspecification for the random effects will also be investigated in Section \ref{sec:simulations}.
%

\subsection{RDS weights}\label{sec:RDSweights}
When conventional sampling methods are used to gather information on a target population, sampling probabilities are known throughout the sampling process. This allows the researcher to compute and take into account design weights when estimating finite population parameters. These approaches are infeasible in an RDS setting since sampling probabilities are unknown. The sampling process is only (partially) controlled by the researcher through the selection of an initial set of seeds -- who, while carefully chosen, still represent a convenience sample -- with the remainder of the recruitment working through a sampling mechanism that relies on individuals' social networks and personal decisions. Let $R_{ij}=1$ if the $j$th individual in the $i$th cluster is sampled. If the true sampling design $\mathcal{S}$ were known, the inclusion probability of the $j$th individual in the $i$th cluster would be computed as
$$
\pi_{ij}=\mbox{E}(R_{ij}|\mathcal{S}).
$$
\cite{Vol08} approximated the RDS process as a random walk on the nodes of an undirected graph and treated RDS samples as independent draws from its stationary distribution. The resulting inclusion probability for the $j$th individual is estimated by
$$
\hat{\pi}^{RDS-II}_{ij}=\frac{1}{n_{ij}}\frac{\sum_{i=1}^{m}\sum_{j=1}^{n_i}n_{ij}}{n}.
$$
Recalling that  $n_{ij}$ is the number of social connections that the $j$th individual of the $i$th cluster shares with others in the same cluster, these weights have the appealing intuition of adjusting for the `popularity' of an individual, and hence their likelihood of being recruited. 
\cite{Gile11} showed that the resulting estimators for means and proportions can be severely biased when sample fractions are large, among other factors. They proposed successive sampling (SS) weights based on a SS approximation of the RDS sampling design, which is viewed as a probability proportional to size without replacement design, and showed that resulting estimators consistently outperform estimators based on RDS-II weights.
Details of the algorithm for computing the weights are given in \cite{Gile11}. An important drawback of this approach is that the computation of inclusion probabilities requires knowledge of the population size. Another is that the weights vary depending on the chosen outcome, and so must be computed a new for each outcome or analysis; this can be impractical in large, collaborative or multi-site studies.
%

Until recently, the majority of inferential methods in the RDS literature dealt with the estimation of population means or proportions. The use of RDS weights in these settings is principled and straightforward. 
The use of sampling weights in a regression setting is more challenging (see, for example, \citealt{lohr2009sampling} for a more thorough and rigorous discussion of these issues in general, and \cite{Spi09} for a discussion regarding RDS regression in particular).
 In light of these discussions, 
we consider the use of unit-level weights -- specifically RDS-II and SS weights -- when fitting the model as a way of taking the RDS design information into account, as these are widely used in the RDS literature.

\subsection{Bootstrap variance estimators}
We consider two bootstrap methods for estimating uncertainty in RDS: $(i)$ the \textit{tree} bootstrap \citep{baraff2016estimating} and $(ii)$ the \textit{neighborhood} bootstrap \citep{2020arXiv201000165Y}.

The tree bootstrap method is based on resampling the RDS tree. Bootstrap samples are typically drawn from the observed recruitment tree by mimicking its hierarchical structure. The first level of the tree generation consists of resampling with (or without) replacement from the sets of seeds of the observed recruitment tree. In the second level of the bootstrap procedure, we resample with (or without) replacement from each of the sampled seeds' recruits. The third level is created by resampling from the wave 1 participants' recruits. The process continues until there are no more recruits from which to sample. The tree bootstrap method mimics the recruitment tree and corresponding features such as the recruitment chain, the number of seeds and waves, thus taking into account the underlying network structure of RDS. Recent findings suggest that this method consistently outperforms existing bootstrap methods, but overestimates uncertainty (\citealt{baraff2016estimating,Gile18,2020arXiv201000165Y}).

The neighborhood bootstrap method is based on sequentially resampling individuals and their neighbors within the RDS tree. The first stage of resampling consists of uniformly selecting $n/c_r$ recruits, where $c_r$ is the average number of connections within the resampled RDS tree. We then include, in the second stage of resampling, the neighbors of all selected recruits in the bootstrap sample. The network component of the resampled RDS data is the subgraph induced by the selected recruits and their neighbors. This method captures the `local' neighborhood structure of the network by reporting all connections that a resampled unit has within the tree, without much reliance on its branching structure. \cite{2020arXiv201000165Y} empirically showed that the neighborhood bootstrap outperforms the tree bootstrap in terms of coverage, bias and mean interval width under realistic RDS assumptions.

\section{Simulations}\label{sec:simulations}
We conducted two separate simulation studies to assess the accuracy of regression parameter estimators under two distinct modeling scenarios. Under the assumption that Equation (\ref{GLMmodel}) is the data-generating model, and that the variable $x$ is uncorrelated with degree, the goal of the first simulation study is to assess the accuracy, precision and coverage of the 95\% (model-based and bootstrap) confidence intervals for estimators of $\beta_1$ if $(i)$ homophily-driven effects $\gamma$ are ignored when present and $(ii)$ the correlation model (\ref{SAR}) for the random effects $\bm{\delta}_i$ is misspecified. We consider fitting the model without RDS weights, with RDS-II weights, and with SS weights under three potential population sizes (one of which is correct). 
In the second simulation study, we assume a simpler version of the data-generating model (\ref{GLMmodel}) with no homophily-driven effects (implying that there are no missing covariates in the subsequent fitted model) and assess the accuracy, precision and coverage of the 95\% confidence intervals for estimators of $\beta_1$ when the variable $x$ is correlated with degree.

\subsection{RDS sampling}\label{sec:paramsamp}
We simulated networks using Exponential Random Graph Models (ERGM) \citep{harris2014introduction}, a class of generative models for modeling network dependence. Let $\mathbf{S}$ be the random adjacency matrix of the network, and $\mathbf{x}$ a vector of nodal attributes. The joint distribution of its elements is:
\begin{equation}\label{ergm}
\mbox{P}\left(\mathbf{S}=\mathbf{s}|\mathbf{x},\bm{\eta}\right)=\frac{\mbox{exp}\left\lbrace \bm{\eta} g\left(\mathbf{s},\mathbf{x}\right) \right\rbrace }{\bm{\kappa\left(\bm{\eta}\right)}},
\end{equation}
where $\bm{\eta}$ is a vector of parameters and $g\left(\mathbf{a},\mathbf{x}\right) $ its corresponding vector of network statistics, $\bm{\kappa\left(\bm{\eta}\right)}=\sum_{\mathbf{s}} \mbox{exp}\left\lbrace \bm{\eta} g\left(\mathbf{s},\mathbf{x}\right) \right\rbrace$ is a normalizing constant. The features of the network are captured in (\ref{ergm}) by choosing network statistics to represent density ($\mathcal{D}_n$) or the ratio of ties in the observed network over the total number of possible ties, degree distribution and homophily. The degree distribution is mainly controlled by setting different values for the Geometrically-Weighted Degree parameter $\eta_{G}$ along with a `decay' parameter $\eta_{d}$ that controls for the level of geometric weighting. When $\eta_{G}<0$ there are more high- and low-degree individuals than expected by chance, while when $\eta_{G}>0$ the network is more centralized \citep{Lev16}.
We simulated 10 clusters of equal sizes from which RDS samples were drawn for the following set of network and sample characteristics.
The population size $N=1000$, $\mathcal{D}_n=1\%$, and $\eta_{G}(\eta_d)=-6 (3)$. We consider $s=10$ seeds, $c=3$ coupons, and sampling fractions of either 20\%, 50\%, or 80\%. We also consider RDS-II weights ($\pi_{RDS}$), SS weights with $N$ known ($\pi_{SS}$), SS weights with $\hat N_u=N-(N-n)/2$ ($\pi_{SS}^u$) and SS weights with $\hat N_o=N+(N-n)/2$ ($\pi_{SS}^o$).

\subsection{Regression models}\label{sec:regsimresults}
	In the first simulation study, we generated a continuous covariate $X$ from a normal distribution with mean $3$ and standard deviation $1.5$. We define the following model:
$$
g(\mu_{ij})=\beta_0+\beta_1 X_{ij}+\gamma \frac{1}{n_{ij}}\sum_{k\sim j}X_{ik}+\delta_{ij},
$$
where $\delta_{ij}$ follows the SAR model (\ref{Sigmaneighbour}). We set the parameter vector to $(\beta_0,\beta_1, \gamma, \sigma^2)=(0,\,2,\,1.5,\,1)$ for each value of the autocorrelation parameter $\rho=0.05,\,0.1$.
We considered three link functions: $g(\mu_{ij})=\mu_{ij}$, $g(\mu_{ij})=\log(\mu_{ij})$ and $g(\mu_{ij})=\mbox{logit}(\mu_{ij})$; for the logistic model, we set the prevalence of the outcome variable to $30\%$ by calibrating the intercept parameter to $\beta_0=-12$ using the cumulative distribution function of the logistic distribution.  For each combination of network and sample characteristics, we fitted models in which the parameter $\gamma$ is ignored.

	In the second simulation study, we assume the following data-generating model:
$$
g(\mu_{ij})=\beta_0+\beta_1 X_{ij}+\delta_{ij},
$$
where $\delta_{ij}$ follows the SAR model (\ref{Sigmaneighbour}). The parameter vector is set to $(\beta_0,\beta_1, \sigma^2, \rho)=(0,\,2,\,1,\,0.05)$. We generated the continuous covariate $X$ in such a way that the correlation with degree, measured using the Pearson correlation coefficient, is $\rho_d=0.4$ or $0.6$. The setting for the link functions, the population network and the RDS process are identical to that of the first simulation study; the sample fraction is fixed at $20\%$ across all combinations of simulation parameters.

To account for the dependence between observations in models for both simulations, we assumed clustering at both seed and recruiter levels, with seed-specific and recruiter-specific random effects. 
 We weighted the models using the set of RDS weights described in Section \ref{sec:paramsamp}; we assumed that each individual's reported network size is precisely known. RDS-II and SS weights were computed via \textsf{vh.weights} and \textsf{gile.ss.weights} respectively, both functions of the \textsf{R} package \textsf{RDS}. We computed the relative bias and the root mean squared error of $\hat{\beta}_1$, and the coverage of the $95\%$ (model-based and bootstrap) confidence intervals for ${\beta_1}$.

\subsection{Results from the first simulation study: ignoring homophily-driven effects and/or misspecifying the correlation model}

	Tables \ref{table:simulation_Lin}, \ref{table:simulation_P} and \ref{table:simulation_Lk} report the relative bias and the root mean squared error of $\hat \beta_1$, and the coverage of the $95\%$ confidence interval for $\beta_1$ in the linear, Poisson and logistic regression cases, respectively. Additional results for a smaller sample fraction ($f=10\%$) are reported in tables S1-S3 of the Web supplement.

	For the linear regression, estimators are unbiased across all sampling fractions and network dependence parameters considered. The precision minimally increases with increasing sample fractions, but decreases with increasing network dependence. The coverage of the $95\%$  confidence interval is consistently close to the nominal value; the unweighted estimator offers better coverage than weighted estimators.
	
	For the Poisson regression, estimators exhibit small biases across all sample fractions and network dependence; the unweighted estimator is slightly less biased than weighted estimators. The bias slightly increases with an increasing network dependence but does not consistently decrease with an increasing sample fraction. As in the linear case, the precision minimally increases with an increasing sample size, but does not consistently decrease with an increasing network dependence. The coverage of the $95\%$ model-based confidence intervals are far below their nominal values; the coverage for the tree bootstrap confidence interval exceeds or is at the nominal value while, for the neighborhood confidence interval, the coverage is slightly below or at the nominal value. Assuming clustering at the recruiter level offers better coverage than assuming clustering at the seed level.
	
	The logistic regression analysis yields estimators that are heavily biased across all sampling fractions, network dependence and sampling weights when clustering is assumed at the seed level. Models that assume clustering at the recruiter level yield estimators that exhibit small to negligible biases. The coverage of the model-based confidence intervals are below their nominal values; the coverage for the tree bootstrap confidence interval is above the nominal value, and the coverage for the neighborhood bootstrap is slightly below or at the nominal value in most cases, when the bias is small to negligible. Again, clustering on the recruiter level yields better coverage.

	These results are consistent with previous findings that omitting a non-confounding covariate (assuming the random effects model is correctly specified) does not induce bias for the linear and the Poisson regressions. In the logistic regression case, the omission of the covariate for the homohpily-driven effects induces attenuation bias because of the inappropriate collapsing of the contingency tables (\citealt{Cologne2019EffectsOO,Gail1984BiasedEO}).

	To better understand the observed coverage for the Poisson and logistic regressions, we reported the relative biases for the model-based and the bootstrap variance estimators in Web Supplement tables S4-S6. 
The model-based variance estimator underestimates uncertainty across all sampling fractions, levels of clustering and network dependence. The tree bootstrap variance estimator severely overestimates uncertainty in most cases while the neighborhood bootstrap variance estimator is, in absolute value, less biased than both estimators in most cases, especially for the linear model. This aligns with previous findings in the RDS literature that, for the tree bootstrap method, covering at or above the nominal level generally comes at a significant cost in terms of power (\citealt{Gile18,2020arXiv201000165Y}).

\begin{table}[H]
\caption{\small Linear -  Relative bias and root mean squared error of $\hat \beta_1$ , model-based coverage (CI), the tree bootstrap coverage (TCI) and the neighborhood bootstrap coverage (NCI) of the $95\%$ confidence interval of $\beta_1$ for increasing levels of sample fraction ($f$), network dependence ($\rho$) and various RDS weights ($\pi$). Clustering (Clstr.) is assumed at the seed level (S) and at the recruiter level (R).}
\begin{center}
  \setlength\extrarowheight{-3pt}
\footnotesize
\begin{tabular}{llc ccccc c ccccc} \hline
\multicolumn{6}{r}{$f=20\%$}&&&&&&{$f=80\%$}\\
    \cline{4-8}     \cline{10-14}
{$\rho$}& {Clstr.} & {$\pi$}& {RB}&{RMSE}&{CI}&{TCI}&{NCI}& & {RB}&{RMSE}&{CI}&{TCI}&{NCI}\\ \hline
\multirow{5}{1em}{$.05$}& \multirow{1}{1em}{S}   & 1 & 0 &0.06  &0.96  &  0.99&0.94&&0 &0.03&0.94 &1.00&0.92 \\
&  &  $\pi_{RDS}$ &0 & 0.07 & 0.90  & 0.98&0.93& &0&0.05&0.82&1.00&0.92 \\
 & &$\pi_{SS}$ &0  &  0.07   & 0.92   &0.98 &0.93  &&0 &0.04&0.90& 1.00&0.91\\
   & &$\pi_{SS}^u$ &0 & 0.07 & 0.93   &0.98 & 0.94 &&0&0.04&0.91& 1.00&0.91\\
 & &$\pi_{SS}^o$  & 0  &  0.07    &0.92   & 0.98 &0.93 &&0 &0.04&0.89&1.00 &0.91 \\ \\
 & \multirow{1}{1em}{R}   & 1 & 0 &0.07  &0.95   &0.99& 0.94& &0 &0.03&0.94&1.00& 0.95 \\
&  &  $\pi_{RDS}$ &0 & 0.08 & 0.88  & 0.99&0.93 &&0&0.05&0.81&1.00& 0.95\\
 & &$\pi_{SS}$ &0  &  0.07   & 0.89   &0.99&0.93&  &0 &0.04&0.91& 1.00&0.94\\
   & &$\pi_{SS}^u$ &0 & 0.07 & 0.90   & 0.99 &0.94 &&0&0.04&0.92&0.99&0.94 \\
 & &$\pi_{SS}^o$  & 0  &  0.08    &0.88   & 0.99& 0.93&&0 &0.04&0.90& 0.99 &0.93\\
\hline
\multirow{5}{1em}{$.1$} & \multirow{1}{1em}{S}   &1 &0  &0.07 &0.96 &1.00&0.94&  &0&0.04& 0.94&1.00& 0.96\\
   &  &$\pi_{RDS}$ & 0 & 0.09& 0.91  &0.98&0.94& &0&0.06&0.84&0.99 &0.95\\
 & &$\pi_{SS}$ & 0 & 0.08& 0.94   &0.99 & 0.95&&0&0.04&0.93&0.99&0.95 \\
  & &$\pi_{SS}^u$  & 0&0.08 &  0.94  &0.99&0.95 &&0&0.04&0.93&1.00&0.95 \\
 &  & $\pi_{SS}^o$ & 0&0.08& 0.93  &  0.99&0.94 & &0&0.04&0.92& 0.99&0.96\\ \\
 & \multirow{1}{1em}{R}   & 1 &0  &0.08 &0.93 &1.00&0.94&  &0&0.04& 0.93&1.00 &0.96\\
   &  &$\pi_{RDS}$ & 0 & 0.09& 0.89  &1.00 &0.93 &&0&0.06&0.82& 1.00&0.96\\
 & &$\pi_{SS}$ & 0 & 0.09& 0.89   &  1.00& 0.92& &0&0.04&0.91& 1.00&0.95\\
  & &$\pi_{SS}^u$  & 0&0.09 &  0.91  & 1.00& 0.92 &&0&0.04&0.91&1.00 &0.95\\
 &  & $\pi_{SS}^o$ & 0&0.09& 0.89  & 1.00&0.93 &&0&0.04&0.91&1.00& 0.96\\
\hline
\end{tabular}
\end{center}
\label{table:simulation_Lin}
\end{table}

\begin{table}[H]
\caption{\small Poisson - Relative bias and root mean squared error of $\hat \beta_1$ , model-based coverage (CI), the tree bootstrap coverage (TCI) and the neighborhood bootstrap coverage (NCI) of the $95\%$ confidence interval of $\beta_1$ for increasing levels of sample fraction ($f$), network dependence ($\rho$) and various RDS weights ($\pi$). Clustering (Clstr.) is assumed at the seed level (S) and at the recruiter level (R).}
\begin{center}
  \setlength\extrarowheight{-3pt}
\footnotesize
\begin{tabular}{llc ccccc c ccccc} \hline
\multicolumn{6}{r}{$f=20\%$}&&&&&&{$f=80\%$}\\
    \cline{4-8}     \cline{10-14}
{$\rho$}& {Clstr.} & {$\pi$}& {RB}&{RMSE}&{CI}&{TCI}&{NCI}& & {RB}&{RMSE}&{CI}&{TCI}&{NCI}\\ \hline
\multirow{5}{1em}{$.05$}  & \multirow{1}{1em}{S}  & 1 & -0.05 & 0.41 & 0.41 & 0.98&0.90 & &-0.08&0.30&0.29 &0.87&0.80\\
& &    $\pi_{RDS}$ & -0.08 &  0.49& 0.38  &0.97&0.89&&-0.12&0.38&0.23&0.82&0.79\\
 && $\pi_{SS}$ & -0.08 &  0.48   &  0.39  &0.97&0.89& &-0.10&0.34&0.24&0.86&0.80\\
   && $\pi_{SS}^u$ & -0.08 &  0.46&  0.38  &0.98&0.89 &&-0.10&0.33&0.25&0.85&0.80\\
 && $\pi_{SS}^o$  & -0.08  &  0.48    &   0.40 &0.97 &0.89&&-0.10&0.34&0.24&0.85 &0.79\\ \\
 & \multirow{1}{1em}{R}  & 1 & -0.05 & 0.36 & 0.51   &0.96&0.93 &&-0.06&0.24&0.36& 0.97&0.92\\
& &    $\pi_{RDS}$ & -0.06 &  0.46& 0.46  & 0.93&0.93&&-0.09&0.30&0.31&0.96&0.94\\
 && $\pi_{SS}$ & -0.08 &  0.47   &  0.53  &0.94 &0.93&&-0.08&0.27&0.36&0.97&0.94\\
   && $\pi_{SS}^u$ & -0.08 &  0.45&  0.52  & 0.95 &0.93 &&-0.08&0.27&0.31&0.97&0.94\\
 && $\pi_{SS}^o$  & -0.08  &  0.47  &   0.55 & 0.93&0.92&&-0.08&0.28&0.36&0.97 &0.92\\
\hline
\multirow{5}{1em}{$.1$}& \multirow{1}{1em}{S}    & 1 &-0.08  & 0.43 &0.42   &0.92&0.90& &-0.12&0.35&0.27&0.96&0.79 \\
&   & $\pi_{RDS}$ & -0.10  &0.44  &0.37   & 0.90&0.88&&-0.15&0.40&0.23&0.95&0.80\\
& & $\pi_{SS}$ &-0.10   & 0.44& 0.39   & 0.91&0.90&&-0.14&0.37&0.24&0.95&0.79\\
&  & $\pi_{SS}^u$  & -0.10 &0.44 &  0.40 &0.91 &0.90& &-0.13&0.37&0.24&0.95&0.78\\
 & &  $\pi_{SS}^o$ &-0.10  & 0.44&  0.39  & 0.91 &0.90&&-0.14&0.38&0.24&0.95&0.80\\\\
 & \multirow{1}{1em}{R}    & 1 &-0.10  & 0.52 &0.47 &0.97&0.96& &-0.09&0.31&0.32& 0.98&0.94\\
&   & $\pi_{RDS}$ & -0.11 &0.51  &0.48   &  0.95&0.96&&-0.11&0.34&0.29&0.98&0.94\\
& & $\pi_{SS}$ &-0.13  & 0.63& 0.56  &0.95&0.94& &-0.10&0.33&0.32&0.99&0.94\\
&  & $\pi_{SS}^u$  & -0.12 &0.60 &  0.54 &0.95  &0.94 & &-0.10&0.33&0.33&0.99&0.96\\
 & &  $\pi_{SS}^o$ &-0.13  & 0.63&  0.60 & 0.94& 0.92&&-0.10&0.34&0.33&1.00&0.96\\
\hline
\end{tabular}
\end{center}
\label{table:simulation_P}
\end{table}

\begin{table}[H]
\caption{\small Logistic - Relative bias and root mean squared error of $\hat \beta_1$ , model-based coverage (CI), the tree bootstrap coverage (TCI) and the neighborhood bootstrap coverage (NCI) of the $95\%$ confidence interval of $\beta_1$ for increasing levels of sample fraction ($f$), network dependence ($\rho$) and various RDS weights ($\pi$). Clustering (Clstr.) is assumed at the seed level (S) and at the recruiter level (R).}
\begin{center}
  \setlength\extrarowheight{-3pt}
\footnotesize
\begin{tabular}{llc ccccc c ccccc} \hline
\multicolumn{6}{r}{$f=20\%$}&&&&&&{$f=80\%$}\\
    \cline{4-8}     \cline{10-14}
{$\rho$}& {Clstr.} & {$\pi$}& {RB}&{RMSE}&{CI}&{TCI}&{NCI}& & {RB}&{RMSE}&{CI}&{TCI}&{NCI}\\ \hline
\multirow{5}{1em}{$.05$}& \multirow{1}{1em}{S}   & 1 & -0.17 & 0.43 & 0.71&  0.92 &0.76& &-0.21&0.43&0.43&0.59&0.45 \\
&&    $\pi_{RDS}$ & -0.15 &  0.47& 0.62& 0.92 &0.84&&-0.23&0.49&0.19&0.73&0.42\\
 && $\pi_{SS}$ & -0.16 &  0.46   &  0.63&0.93 &0.82& &-0.22&0.46&0.34&0.57&0.38\\
  & & $\pi_{SS}^u$ & -0.16 &  0.46&  0.64& 0.92 &0.80& &-0.22&0.45&0.37&0.55&0.39\\
 && $\pi_{SS}^o$  & -0.16  &  0.47    &   0.63&0.93 &0.83& &-0.22&0.47&0.31&0.59& 0.38\\ \\
 & \multirow{1}{1em}{R}   & 1 & 0.12 & 1.35 & 0.63 &  1.00&0.98& &0.01&0.43&0.54&1.00& 0.98\\
&&    $\pi_{RDS}$ & 0.15 &  1.19& 0.63&  1.00&0.99&&0.03&0.36&0.61&1.00&0.97\\
 && $\pi_{SS}$ & 0.14 &  1.20   &  0.62 & 1.00&0.99 &&0.03&0.39&0.54&1.00&0.98\\
  & & $\pi_{SS}^u$ & 0.14 &  1.27&  0.62& 1.00&0.99&&0.03&0.40&0.53&1.00&0.98\\
 && $\pi_{SS}^o$  & 0.14 &  1.19    &   0.63& 1.00&0.99& &0.03&0.38&0.57 &1.00&0.98\\
\hline
\multirow{5}{1em}{$.1$} & \multirow{1}{1em}{S}   & 1 &-0.24  & 0.53 &0.61& 0.85 & 0.74& &-0.25&0.51&0.35&0.30&0.20 \\
  & & $\pi_{RDS}$ & -0.21  &0.52  &0.53&  0.95 &0.82& &-0.25&0.53&0.18&0.59&0.23\\
& & $\pi_{SS}$ &-0.22  & 0.52& 0.54& 0.95&0.82& &-0.25&0.52&0.30&0.32&0.25\\
&  & $\pi_{SS}^u$  & -0.22 &0.52 &  0.52& 0.94&0.80&  &-0.25&0.52&0.34&0.30&0.25\\
& &  $\pi_{SS}^o$ &-0.21 & 0.52&  0.53 & 0.95&0.82& &-0.25&0.52&0.28&0.37&0.25\\ \\
& \multirow{1}{1em}{R}   & 1 &-0.01  & 0.83 &0.58&  1.00 &0.98& &-0.08&0.36&0.50&1.00 &0.98\\
  & & $\pi_{RDS}$ & 0.05  &0.80  &0.65&  1.00 & 0.99&&-0.04&0.31&0.59&1.00&0.99\\
& & $\pi_{SS}$ &0.06  & 1.15& 0.62&1.00&0.99& &-0.06&0.32&0.49&1.00&0.99\\
&  & $\pi_{SS}^u$  & 0.05 &1.11 &  0.60 &1.00&0.99&  &-0.07&0.33&0.49&1.00&0.98\\
& &  $\pi_{SS}^o$ &0.07& 1.17&  0.63  &1.00&0.99 &&-0.06&0.31&0.51&1.00&0.99\\
\hline
\end{tabular}
\end{center}
\label{table:simulation_Lk}
\end{table}

\subsection{Results for the second simulation study: correlated predictor and degree}
Tables \ref{table:simulation_Lin_corr}, \ref{table:simulation_P_corr} and \ref{table:simulation_Lk_corr} report the results for the linear, Poisson and logistic regression respectively.  Weighted estimators are (slightly) less biased than unweighted estimators across all models, clustering levels and levels of correlation between the predictor and the degree. Further, RDS-II weights perform as well as the SS weights across all models.

\begin{table}[H]
\caption{ \small Linear, with degree/covariate correlation - Relative bias and root mean squared error of $\hat \beta_1$, model-based coverage (CI), tree bootstrap coverage (TCI) and neighborhood bootstrap coverage (NCI) of the $95\%$ confidence interval of $\beta_1$ with increasing association between predictor and degree ($\rho_d$) and for various RDS weights ($\pi$).  Clustering (Clstr.) is assumed at the seed level (S) and at the recruiter level (R).}
\begin{center}
\setlength\extrarowheight{-3pt}
\footnotesize
\begin{tabular}{lc ccccc c ccccc} \hline
\multicolumn{5}{r}{$\rho_d=0.4$}&&&&&&{$\rho_d=0.6$}\\
    \cline{3-7}     \cline{9-13}
 {Clstr.} & {$\pi$}& {RB}&{RMSE}&{CI}&{TCI}&{NCI}& & {RB}&{RMSE}&{CI}&{TCI}&{NCI}\\ \hline
 \multirow{1}{1em}{S}   & 1 & 0 &0.02 &0.98  &  0.99&0.95&& 0 &0.02 &0.95&0.99&0.93 \\
  &  $\pi_{RDS}$ &0 & 0.02 & 0.84  & 0.99&0.91 &  &0&0.02&0.89&0.97 &0.90\\
  &$\pi_{SS}$ &0  &  0.02   & 0.85  &0.99 &0.91& &0&0.02&0.90&0.97&0.91\\
    &$\pi_{SS}^u$ &0 & 0.02 & 0.85   &0.99 & 0.91 &  &0 &0.02&0.90&0.98&0.91\\
  &$\pi_{SS}^o$  & 0  &  0.02    &0.85   & 0.99&0.91 & &0 &0.02&0.90&0.97&0.91 \\ \\
  \multirow{1}{1em}{R}   & 1 & 0 &0.01  &0.96   &0.99&0.95& & 0 &0.02 &0.95&0.99 & 0.92\\
  &  $\pi_{RDS}$ &0 & 0.02 & 0.88  & 0.99& 0.93 & &0&0.02&0.86&0.99 &0.91\\
  &$\pi_{SS}$ &0  &  0.02   & 0.90   &0.99& 0.93& &0&0.02&0.87&0.99&0.92\\
    &$\pi_{SS}^u$ &0 & 0.02 & 0.92   & 0.99 &0.93 &  &0 &0.02&0.88&0.99& 0.91\\
  &$\pi_{SS}^o$  & 0  &  0.02 &0.89   & 0.99& 0.93&  &0 &0.02&0.87&0.99&0.91\\
\hline
\end{tabular}
\end{center}
\label{table:simulation_Lin_corr}
\end{table}

\begin{table}[H]
\caption{\small Poisson, with degree/covariate correlation - Relative bias and root mean squared error of $\hat \beta_1$, model-based coverage (CI), tree bootstrap coverage (TCI) and neighborhood bootstrap coverage (NCI) of the $95\%$ confidence interval of $\beta_1$ with increasing association between predictor and degree ($\rho_d$) and for various RDS weights ($\pi$).  Clustering (Clstr.) is assumed at the seed level (S) and at the recruiter level (R).}
\begin{center}
\setlength\extrarowheight{-3pt}
\footnotesize
\begin{tabular}{lc ccccc c ccccc} \hline
\multicolumn{5}{r}{$\rho_d=0.4$}&&&&&&{$\rho_d=0.6$}\\
    \cline{3-7}     \cline{9-13}
 {Clstr.} & {$\pi$}& {RB}&{RMSE}&{CI}&{TCI}&{NCI}& & {RB}&{RMSE}&{CI}&{TCI}&{NCI}\\ \hline
 \multirow{1}{1em}{S}  & 1 & -0.02 & 1.01& 0.32 & 0.94 &0.92&&-0.01&1.26&0.29&0.97&0.93\\
&    $\pi_{RDS}$ & -0.01 &  0.82& 0.33  &0.96& 0.92&  & 0&1.02&0.32&0.96&0.92\\
 & $\pi_{SS}$ & -0.01 &  0.83  &  0.32  &0.96&0.92& & 0&1.06&0.31&0.96&0.92\\
   & $\pi_{SS}^u$ & -0.01 &  0.87&  0.32  &0.95& 0.92 &  & -0.01&1.11&0.31&0.96&0.92\\
 & $\pi_{SS}^o$  & -0.01  &  0.85    &   0.32&0.96 &0.92 & &  0 &1.05&0.32&0.96&0.92\\ \\
  \multirow{1}{1em}{R}  & 1 & -0.05 & 2.17& 0.35 &0.98&0.96&&-0.03&0.30 &0.55&0.97&0.95\\
 &    $\pi_{RDS}$ & -0.02 & 1.04& 0.38  & 0.98&0.95&  & -0.02&0.24&0.60&0.98&0.96\\
 & $\pi_{SS}$ & -0.03 & 1.26 &  0.37  &0.98 &0.95&& -0.02&0.25&0.60& 0.98&0.96\\
   & $\pi_{SS}^u$ & -0.03 &  1.09&  0.35 & 0.98 &0.95 & & -0.02&0.26&0.60&0.98&0.96\\
 & $\pi_{SS}^o$  & -0.04  &  1.59&   0.37& 0.98&0.95  &&  -0.02 &0.25&0.60& 0.98&0.96\\
\hline
\end{tabular}
\end{center}
\label{table:simulation_P_corr}
\end{table}

\begin{table}[H]
\caption{\small Logistic, with degree/covariate correlation - Relative bias and root mean squared error of $\hat \beta_1$, model-based coverage (CI), tree bootstrap coverage (TCI) and neighborhood bootstrap coverage (NCI) of the $95\%$ confidence interval of $\beta_1$ with increasing association between predictor and degree ($\rho_d$) and for various RDS weights ($\pi$).  Clustering (Clstr.) is assumed at the seed level (S) and at the recruiter level (R).}
\begin{center}
\setlength\extrarowheight{-3pt}
\footnotesize
\begin{tabular}{lc ccccc c ccccc} \hline
\multicolumn{5}{r}{$\rho_d=0.4$}&&&&&&{$\rho_d=0.6$}\\
    \cline{3-7}     \cline{9-13}
 {Clstr.} & {$\pi$}& {RB}&{RMSE}&{CI}&{TCI}&{NCI}& & {RB}&{RMSE}&{CI}&{TCI}&{NCI}\\ \hline
 \multirow{1}{1em}{S}   & 1 & -0.14 & 0.39 & 0.82&  0.95 &0.80&&-0.14&0.37 &0.79&0.94 &0.82\\
&    $\pi_{RDS}$ & -0.11 &  0.41& 0.84& 0.99 &0.88&   & -0.10&0.37&0.79&0.97&0.91\\
 & $\pi_{SS}$ & -0.11 &  0.40   &  0.83&0.99 & 0.87&& -0.11&0.37&0.76&0.97&0.90\\
   & $\pi_{SS}^u$ & -0.12 &  0.39&  0.84& 0.99 &  0.86 & & -0.11&0.37&0.76&0.97&0.89\\
 & $\pi_{SS}^o$  & -0.11  &  0.40    &   0.85&0.99 & 0.87 & &  -0.11 &0.37&0.78&0.97& 0.91\\ \\
  \multirow{1}{1em}{R}   & 1 & -0.05 & 0.52 & 0.46 & 1.00&0.98&&-0.05&0.39 &0.88&1.00 &0.98\\
&    $\pi_{RDS}$ & 0.04 &  0.54& 0.62& 1.00&0.99 &  & 0.04&0.44&0.87&1.00&0.98\\
 & $\pi_{SS}$ & 0.03 &  0.52   &  0.60& 1.00&0.98& & 0.03&0.43&0.88&1.00&0.98\\
   & $\pi_{SS}^u$ & 0.01 &  0.50&  0.54&1.00& 0.98  & & 0.02&0.43&0.86&1.00&0.98\\
 & $\pi_{SS}^o$  & 0.03 &  0.52    &   0.61& 1.00&  0.98 &&  0.03 &0.43&0.87&1.00&0.98\\
\hline
\end{tabular}
\end{center}
\label{table:simulation_Lk_corr}
\end{table}

\subsection{Summary and guidelines}

Our results show that ignoring homophily-driven effects, if present, induces a negligible to small bias for linear and Poisson models while, for the logistic regression, this strategy induces a substantial bias in the estimates when clustering is assumed at the seed level, and less bias but increased variability when clustering is assumed at the recruiter level. Moreover, misspecifying the SAR correlation model for the random effects induces an increasing bias as the dependence within the network increases, as well as a poor coverage of the model-based confidence interval for the Poisson and logistic regressions. Bootstrap-based confidence intervals yield better coverage than model-based confidence intervals, particularly for Poisson and logistic regressions. Also, fitting mixed models in which clustering is assumed at the recruiter level yields estimators with less bias and better coverage than models in which clustering is assumed at the seed level.
	
	As for RDS weights, unweighted regression methods consistently outperform weighted methods in terms of precision and coverage when the predictor is uncorrelated with degree at the population level. The difference in precision can be attributed to the diffusion of the degree distribution of the network, which resulted in individuals having more small and large weights than expected by chance, hence increased variability in the estimates \citep{Ave19}. On the other hand, weighted regression methods consistently outperform unweighted methods in terms of bias when the predictor is correlated with degree.

We can therefore provide some general guidance for regression in RDS studies: $(i)$ analyses that omit homophily-driven effects terms, while including a random effect for recruiter, outperform other modeling strategies in terms of bias and coverage of the confidence interval, and $(ii)$ weighted regression methods outperform unweighted regression methods in terms of bias and precision when the predictor is correlated with degree; when the predictor is uncorrelated with degree, weighting the model only increases variability in the estimates. As observed previously \citep{2020arXiv201000165Y}, neighbourhood bootstrap provides better estimates of standard errors than any existing alternatives.

\section{Case Study}\label{sec: casestudy}
We now turn to an analysis of the Engage study, a Canadian study conducted in three cities: Montreal, Toronto and Vancouver. The study aims to determine the individual, social and community-level risk factors for transmission of HIV and sexually transmitted infections and related behaviours within the GBM community. In this example, we focus on the study conducted in Montreal. The Engage data-analysis team designed two databases and a tracker to monitor the RDS recruitment process. The study led to the recruitment of $n$ = 1179 GBM from Montreal between February 2017 through June 2018. Approximately 45\% of recruited individuals were successful at recruiting, and 82\% of these effective recruiters brought 1 to 3 peers into the study; 6 seeds of a total of 27 seed participants were unsuccessful at starting recruitment chains.

\subsection{Descriptive statistics}
	Treatment optimism was measured on a scale of 12 items, developed by \cite{ven2000scale} to measure attitutes towards HIV treatment within the Australian GBM community. All items were measured on a 4-point Likert scale (strongly disagree, disagree, agree, strongly agree). The optimism score (TMTOPT) was obtained by summing 10 items and substracting 2 items. This gives a range of possible values between 0 (highly skeptical) and 36 (highly optimistic).

Following \cite{levy2017longitudinal} who found age, education and income as correlates of optimism through a range of bivariate analyses, we chose these same socio-demographic characteristics, among others, as possible predictors for treatment optimism. Descriptive (unweighted) statistics for these variables in the sample are presented in Table S7 of the Web Supplement. 
Around $33\%$ of respondents were aged less than 30, about $70\%$ were born in Canada, less than a third had a high school diploma or lower, and about $58\%$ earned less than \$30,000.
Younger and more educated participants are less optimistic with regard to HIV treatment than other socio-demographic groups; the absolute difference is more pronounced for age. Further, participants who were born in Canada and those who earn less than \$30,000 in annual income have higher optimism scores than other participants.

\subsection{Model fitting}
	We chose the potential socio-demographic characteristics correlates 
as predictors of HIV optimism for the aforementioned reasons. We fit various linear mixed-effects models with seed-specific (for comparison purposes) and recruiter-specific random intercepts, in a weighted and unweighted fashion. Parameter estimates, standard errors and $95\%$ (model-based and bootstrap) confidence intervals are reported in Table \ref{table:estimates}.
	
           We performed non-parametric Mann-Whitney U-tests to compare the distribution of degree between groups defined by the socio-demographic characteristics. The null hypothesis of the test is that for randomly selected values of degrees $d_i$ and $d_j$ from two groups, the probability of $d_i$ being greater than $d_j$ is equal to the probability of $d_j$ being greater than $d_i$. In the Engage sample, the p-values of the test for age, education, being born in Canada and annual income are 0.0, 0.13, 0.0 and 0.0, respectively.  This shows differences in the median number of social connections between groups defined by age, being born in Canada and the annual income of participants, thus suggesting the use of weighted regression.

	Guided by the simulations presented in Section \ref{sec:regsimresults} and by the discussion in the preceding paragraph, we focus on the weighted regression estimates with clustering at the recruiter level. We computed standard error estimates and 95\% confidence intervals using the neighborhood bootstrap method. The results show that annual income is significantly (and positively) associated with the optimism about the efficacy of the treatment, with a change of 1.5 points in the expected optimism score.


\begin{table}[H]
\caption{\small Point estimates, standard errors and asymptotic $95\%$ confidence intervals for a linear mixed model applied to the Engage Montreal data, where clustering is assumed at the seed level (S) and at the recruiter level (R), estimated without weights (1), with RDS-II ($\pi_{RDS}$) weights, and SS weights ($\pi_{SS}$). The standard deviation of the random intercept is $\sigma_0$ and the intraclass correlation is $\rho$.}
\begin{center}
  \setlength\extrarowheight{-3pt}
\small
\begin{tabular}{l ccc ccccc} \hline
\multicolumn{4}{r}{S}&&&&{R}\\
    \cline{3-5}     \cline{7-9}
 {$\pi$} & {}& {Est.}&{SE}&{CI}& & {Est.}&{SE}&{CI}\\ \hline
 \multirow{1}{1em}{1}   & Constant & 16.21 &0.48 &[15.3,17.1]&& 16.21 &0.45 &[15.3,17.1]\\
 &  Age ($\leq$ 30) &-0.72 & 0.40 & [-1.5, 0.0] &   &-0.72&0.44&[-1.6, 0.1]\\
  &Education ($<$ college) &0.92  &  0.90  & [-0.8, 2.7]   & &0.85&1.00&[-1.1, 2.8]\\
    &Born in Canada &0.58 & 0.43 & [-0.3, 1.4]   &    &0.55 &0.49&[-0.4, 1.5]\\
     & Annual income ($\leq$ \$30,000) &0.38 & 0.41& [-0.4, 1.2]  &    &0.43 &0.43&[-0.4, 1.3]\\
  &Born in Canada $\times$ Education  & -1.96  & 1.04  &[-4.0, 0.0]   &   &-1.93 &1.15&[-4.2, 0.3]\\
  &$\sigma_0$($\rho$)  & 0.0(0.0)  & -    &-   &   &1.33(0.06)&-&- \\\\
  \multirow{1}{1em}{$\pi_{RDS}$}   & Constant & 15.09 &0.60&[13.9,16.3]   && 15.50 &0.66 &[14.2,16.8]\\
  &  Age ($\leq$ 30) &-0.84 & 0.59 & [-2.0, 0.3] &   &-0.86&0.61&[-2.1, 0.3]\\
  &Education ($<$ college)  &2.62 &  1.42 & [-0.2, 5.4]& &0.59&1.17&[-1.7, 2.9]\\
    &Born in Canada  &0.93 & 0.63 &[-0.3, 2.2] &    &0.56 &0.82&[-1.0, 2.2]\\
     &Annual income ($\leq$ \$30,000) &1.30& 0.69 &[-0.1, 2.6] &    &1.54 &0.62&[0.3, 2.8]\\
  &Born in Canada $\times$ Education &-4.34  &  1.75    &[-7.8, -0.9]&   &-2.53&1.60&[-5.7, 0.6] \\
  &$\sigma_0$($\rho$)  & 0.96(0.01)  & -    &-   &   &3.12(0.15) &-&-\\\\
   \multirow{1}{1em}{$\pi_{SS}$}   & Constant & 15.09 &0.60  &[13.9,16.3] && 15.51 &0.65 &[14.2,16.8] \\
 &  Age ($\leq$ 30) &-0.84 & 0.59 & [-2.0, 0.3]  &   &-0.86&0.60&[-2.0, 0.3]\\
  &Education ($<$ college) &2.62  &  1.41   &[-0.2, 5.4]& &0.63&1.16&[-1.6, 2.9]\\
    &Born in Canada  &0.93 & 0.62 & [-0.3, 2.1]   &    &0.57 &0.80&[-1.0, 2.1]\\
     &Annual income ($\leq$ \$30,000)  &1.29 & 0.68 & [0, 2.6] &    &1.52 &0.61&[0.3, 2.7]\\
  &Born in Canada $\times$ Education  & -4.32  &  1.74   &[-7.7, -0.9] &   &-2.55&1.58&[-5.6, 0.5] \\
  &$\sigma_0$($\rho$)  & 0.95(0.0)  & -    &-   &   &3.10(0.02) &-&- \\
\hline
\end{tabular}
\end{center}
\label{table:estimates}
\end{table}

	It is also worth noting that the directions of the associations between each covariate and the optimism score are consistent across all levels of clustering, regardless of the chosen RDS weight. However, the conclusions in terms of significance of the parameter effects differ whether we fit models with seed-specific random effects or recruiter-specific random effects.
	
	We performed non-parametric hypothesis tests to decide whether or not to weight the model. It is important to highlight that we have not evaluated this approach, but rather use it as an informal tool to guide our analyses. Reasonably, a non significant test does not exclude the possibility that there may be differences in the degree distribution across levels defined by the predictor, suggesting at least the use of weighted regression as a sensitivity check.

	In our analyses, we chose socio-demographic factors as potential predictors of treatment optimism based on available evidence in the literature, but we have not tried to fully understand all predictors of the treatment score construct. Thus, this is a `limited' consideration of all potential predictors of treatment optimism, which can be further extended as more associational studies are conducted on the subject.

\section{Conclusion}\label{sec:discussion}
	The development of regression methods for RDS is limited by a missing data problem as the observed RDS data reveal only partial information about the full, unobserved RDS network. In this case, valid inference about homophily-driven effects and/or network-induced correlation structures cannot be conducted without additional network data or strict topological constraints on the RDS network (Yauck et al., 2020b). We proposed alternative modeling strategies for RDS when the network is partially missing. Our results showed that ignoring homophily-driven effects, if present, induces a small to negligible bias in the parameter estimator (of the homophilic covariate) for linear and Poisson models while inducing a substantial bias for the logistic regression when clustering is assumed at the seed level. Furthermore, misspecifying the correlation model induces an increasing bias as the dependence within the RDS network increases, and poor coverage for the model-based confidence intervals. In this case, the neighborhood bootstrap method yields a variance estimator that is less biased than the model-based and the tree bootstrap variance estimators while offering confidence intervals with coverages that are slightly below 
or at the nominal level for the linear and the Poisson regression. We also showed that weighted regression methods outperform unweighted regression methods in terms of bias when the predictor is correlated with degree, assuming that there is no missing covariate in the model. Weighting the model only adds variability in the estimates when predictor and outcome are uncorrelated.

	In the case study, we restricted our analyses to the Engage Montreal dataset. This could be extended to the analysis of the data collected in Toronto and Vancouver by pooling across cities. This problem of conducting regression analyses using multi-city/state RDS data can be easily embedded within our inferential framework, where we could assume that city-specific networks are drawn from the same population network. This will be the subject of future work.

\bibliographystyle{jasaauthyear}
{\small \bibliography{All-References} }

\end{document}